\pdfoutput=1
\documentclass[
 reprint,
superscriptaddress,
nofootinbib,
 amsmath,amssymb,
 aps,
prl,
]{revtex4-1}
\usepackage{graphicx}
\usepackage{bm}
\usepackage{hyperref}
\usepackage[mathlines]{lineno}

\usepackage[utf8]{inputenc}
\usepackage{amsmath,amssymb}
\usepackage[T1]{fontenc} 
\usepackage{microtype} 
\usepackage[english]{babel} 

\usepackage{booktabs} 

\usepackage{graphicx}
\usepackage{amssymb}
\usepackage{braket,mleftright}
\usepackage{empheq}
\usepackage{subfigure}
\usepackage{stackrel}
\usepackage{soul}
\usepackage{blkarray}
\usepackage{multirow}
\usepackage{amsmath}
\usepackage{physics}
\usepackage{amsfonts}
\usepackage{bm}
\usepackage{bbold} 
\usepackage{color}
\newcommand{\beq}{\begin{equation}}
\newcommand{\eeq}{\end{equation}}
\newcommand{\bse}{\begin{subequations}}
\newcommand{\ese}{\end{subequations}}
\newcommand{\bea}{\begin{eqnarray}}
\newcommand{\eea}{\end{eqnarray}}
\newcommand{\nico}[1]{{\color{black} #1}} 
\newcommand{\nicor}[1]{{\color{black} #1}} 
\newcommand{\nicon}[1]{{\color{black} #1}} 

\newcommand{\nicof}[1]{{\color{black} #1}} 
\usepackage[utf8]{inputenc} 
\usepackage{amsmath}

\usepackage{enumitem} 
\setlist[itemize]{noitemsep} 
\usepackage{hyperref} 
\usepackage{braket}
\usepackage{verbatim} 
\begin{document}


\title{\nicor{Quantum chaos, \nicon{equilibration} and control in extremely short spin chains}}


\author{Nicol\'{a}s Mirkin}
\email[Corresponding author:]{\,mirkin@df.uba.ar}
\affiliation{%
Departamento de F\'{i}sica “J. J. Giambiagi” and IFIBA, FCEyN, Universidad de Buenos Aires, 1428 Buenos Aires, Argentina
}%

\author{Diego Wisniacki}

\affiliation{%
Departamento de F\'{i}sica “J. J. Giambiagi” and IFIBA, FCEyN, Universidad de Buenos Aires, 1428 Buenos Aires, Argentina
}%

\date{\today}%

\begin{abstract}
The environment of an open quantum system is usually modelled as a large many-body quantum system. \nicon{However, when an isolated quantum system itself is a many-body quantum system, the question of how large and complex it must be in order to generate internal equilibration} is an open key-point in the literature. \nicon{In this work, by monitoring the degree of equilibration of a single spin through its purity degradation, we are able to sense the chaotic behaviour of the generic spin chain to which it is coupled. Quite remarkably, this holds even in the case of extremely short spin chains composed of three spins, where we can also reproduce the whole integrable to chaos transition.} Finally, we discuss implications on quantum control experiments and show that quantum chaos reigns over the best degree of control achieved, even in small chains.
\end{abstract}

\maketitle

\textit{Introduction.} 
\nicon{\nicof{Quantum technologies may outperform classical systems for processing information, but this depends on the ability to precisely control a complex many-body quantum system \cite{arute2019quantum,zhong2020quantum}.} \nicof{Furthermore,} since the latter is in general not isolated but in touch with its surrounding, it is also critical to know how to deal with the detrimental effects from the environment. \nicof{As a consequence}, a huge amount of research has been devoted to understand what does \textit{'in touch'} and \textit{'surrounding'} exactly mean in this context \cite{breuer2002theory,zurek2003decoherence,schlosshauer2007decoherence,rivas2012open}.} 

\nicon{A common hypotheses is to consider the environment as a much larger quantum system than the one of interest. However, depending on how well isolated the open quantum system is, \nicof{the time scales introduced by the coupling to the \nicof{outside world} can be much slower than the ones dictating internal equilibration}} \cite{gogolin2016equilibration}. In fact, many of the experiments that are done today consist of working on a few well-isolated qubits and executing controlled operations on some of them \nicof{\cite{dicarlo2009demonstration,blatt2012quantum,gross2017quantum,yoneda2018quantum,neill2018blueprint,landsman2019verified,zinner2016exploring,serwane2011deterministic,murmann2015antiferromagnetic,murmann2015two,walker2018driven}}. In this scenario, one might wonder whether the set of qubits that are not being controlled, by interacting with the qubits that are, may affect  controllability in the same sense that a large environment usually does. \nicon{Therefore, the question of how small and simple this \textit{\nicof{intrinsic environment}} could be to generate \nicof{internal} equilibration and thus affect controllability is absolutely relevant \nicof{\cite{boes2018catalytic,vidiella2014deviations,vidiella2016evolution,deccordi2018simple,dymarsky2018subsystem,bluvstein2020controlling,schiulaz2018few,zisling2020many,wenz2013few}}, not only from a fundamental point of view but also from the experimental side.}

\nicon{\nicof{Unless a correct understanding and efficient characterization of these complex many-body quantum systems is first developed, the ultimate goal of controlling its full dynamics will always remain unattainable.}
\nicof{In this context, great progress has been made in the study of ubiquitous properties} associated with the non-equilibrium dynamics of many-body quantum systems, such as equilibration \cite{linden2009quantum,short2012quantum} and thermalization \nicof{\cite{rigol2008thermalization,rigol2009breakdown,gogolin2016equilibration,neill2016ergodic,kaufman2016quantum,zhu2021observation}}, where quantum chaos plays a major role \cite{deutsch1991quantum,srednicki1994chaos,srednicki1999approach,borgonovi2016quantum}. These works are usually restricted to the limit of high dimensional Hilbert spaces, where the energy spectrum is large enough to assure a proper characterization of quantum chaos through spectral measures. \nicof{It is clear that} this is not possible in the opposite limit, where the many-body quantum system is not sufficiently large.
\nicof{Are there} any vestiges of quantum chaos at this particular limit? The answer to this question is one of the \nicof{main motivations} of our work.}

\nicon{In this Letter, we study to what extent we can extract information about the chaoticity of a large spin \nicof{chain} by sensing a much smaller \nicof{one} with a simple probe. With this purpose, we consider a single spin connected to a generic spin chain and monitor the degree of equilibration of the reduced spin system \nicof{at the limit of infinite temperature} through its purity degradation. Under this framework, we show almost an exact correspondence between the degree of equilibration suffered by the probe and how much chaos is present within the dynamics of the chain, i.e. the more chaos the more equilibration. Quite remarkably, this allows us to reconstruct the whole integrable to chaos transition even in the case of extremely short spin chains composed of three spins. \nicof{The fact of finding} robust vestiges of quantum chaos in such small quantum systems constitutes the main result of our present work.} We believe that the implications of our findings are \nicof{essentially} two. First, since our method does not require \nicof{a diagonalization over huge Hilbert spaces} nor to determine a whole set of symmetrized energy eigenstates \cite{fortes2019gauging, fortes2020signatures,de2020quantum}, it constitutes a novel and easy way of sensing the chaotic behaviour in complicated many-body quantum systems, which may be of experimental interest due to its simplicity \cite{barends2014superconducting,kelly2015state,debnath2016demonstration,li2017measuring, joshi2020quantum}. Second, we argue that this result has relevant implications in quantum control experiments. As we show at the end of our work, the optimal fidelities achieved for a simple control task over the reduced system strongly depend on the chaotic behaviour of the chain. \nicon{In other words, the degree of control is subordinated to the degree of chaos present, even if the spin chain is small.} 

\nico{For concreteness, in the main text we restrict our study to a particular spin chain, but the same analysis can be extended to very different systems, as we show in the Supplemental Material \cite{Note1}.} The system under consideration has no well-defined semiclassical limit and consists on a 1D Ising spin chain with nearest neighbor (NN) interaction and open boundary conditions, described by 
\begin{equation}
H=\sum_{k=1}^{L}\left(h_x \hat{\sigma}_{k}^{x}+h_z\hat{\sigma}_{k}^{z} \right) -   \sum_{k=1}^{L-1}J_{k}\hat{\sigma}_{k}^{z}\hat{\sigma}_{k+1}^{z},  
\label{hamilt}
\end{equation}
where $L$ refers to the total number of spin-$1/2$ sites of the chain, $\hat{\sigma}_{k}^{j}$ to the Pauli operator at site $k=\set{1,2,...,L}$ with direction $j=\set{x,y,z}$, $h_x$ and $h_z$ to the magnetic field in the transverse and parallel direction, respectively, and finally $J_k$ represents the interaction strength within the site $k$ and $k+1$. In general, we will consider equal couplings, i.e. $J_{k}=1 \, \forall \, k=\{1,2...,L-1\}$, situation where the system has a symmetry with respect to the parity operator $\hat{\Pi}$. \nicon{Parity is defined through the permutation operators $\hat{\Pi}=\hat{P}_{0,L-1}\hat{P}_{1,L-2}\dots \hat{P}_{(L-1)/2-1,(L-1)/2+1}$ for a chain of odd length $L$ \nicof{(the even case is analogous).}} This implies that the spanned space is divided into odd and even subspaces with dimension $D=D^{odd}+D^{even}$  ($D^{odd/even}\approx D/2$). However, since in a realistic scenario couplings may be different due to some experimental error, we will also analyze the case with different values for $J_k$ and show the robustness of our result. With respect to the initial conditions, we will consider an initial pure random state as $\ket{\psi(0)}=\ket{\psi_1}\ket{\psi_2}...\ket{\psi_L}$, where each spin at site $k$ initially points in a random direction on its Bloch sphere
\begin{equation}
    \ket{\psi_k}=\cos{\left(\dfrac{\theta_k}{2}\right)}\ket{\uparrow}+e^{i\phi_k}\sin{\left(\dfrac{\theta_k}{2}\right)}\ket{\downarrow},
    \label{initial_state}
\end{equation}
with $\theta_k \in [0,\pi)$ and $\phi_k \in [0,2\pi)$. Note that this ensemble of initial states maximizes the thermodynamic entropy and \nicor{is equivalent} to a situation of infinite temperature \cite{kim2013ballistic}. \nicon{This assumption is  important since the whole spectrum will be equally contributing to the dynamics \cite{Note1}.} From now on, we will take as the reduced system the first spin of the chain and consider the rest as an \textit{effective environment}. For example, a case with $L=3$ represents a single spin acting as an open system and coupled to an effective environment of only two spins. This may sound too simple but we remark that a recent experiment was able to capture chaotic behaviour on a 4-site Ising spin chain by measuring Out-of-Time Ordered Correlators (OTOC's) \cite{fortes2019gauging,PhysRevLett.121.210601} on a nuclear magnetic resonance quantum simulator \cite{li2017measuring}.

In order to fully characterize the integrable to chaos transition, the standard procedure requires the limit of a high dimensional Hilbert space and the separation of the energy levels according to their symmetries \cite{percival1973regular, berry1977level,bohigas1984characterization}. This may demand huge numerical effort or even be quite laborious to implement experimentally. Within all the standard chaos indicators in the literature, in this work we will restrict ourselves to the so-called distribution of $\min(r_n,1/r_n)$, where $r_n$ refers to the ratio between the two nearest neighbour spacings of a given level. By taking $e_n$ as an ordered set of energy levels, we can define the nearest neighbour spacings as $s_n=e_{n+1}-e_n$. With this notation, \nicor{we can measure the presence of chaotic behavior through} \cite{oganesyan2007localization,atas2013distribution,kudo2018finite} 

\begin{equation}
\nicor{\Tilde{r}_n=\dfrac{\min(s_n,s_{n-1})}{\max(s_n,s_{n-1})}=\min(r_n,1/r_n),}
\end{equation}
where $r_n=s_n/s_{n-1}$. As the mean value of $\Tilde{r}_n$ ($\overline{\min(r_n,1/r_n)}$) attains a maximum when the statistics is Wigner-Dyson \nicor{($\mathcal{I}_{WD} \eqsim 0.5307$)} and a minimum when is Poissonian \nicor{($\mathcal{I}_{P} \eqsim 0.386$)}, we can normalize it as
\begin{equation}
    \eta=\dfrac{\overline{\min(r_n,1/r_n)}-\mathcal{I}_P}{\mathcal{I}_{WD}-\mathcal{I}_P}.
\label{eta}
\end{equation}

The parameter $\eta$ quantifies the chaotic behaviour of the system in the sense that $\eta \rightarrow{0}$ refers to an integrable dynamics while $\eta \rightarrow{1}$ to a chaotic one. While $\eta$ is based on the spectral properties of the entire system, \nicon{is useful only in long chains \cite{Note1}} and requires to analyze separately even and odd subspaces, we will show that by studying the \nicor{equilibration} dynamics of a single spin we will be able to reconstruct the whole structure of the regular to chaos transition, even in the case of extremely short spin chains and without resorting to any classification according to the energy level symmetries. 

\begin{figure}[!htb]
\begin{center}
\includegraphics[width=87mm]{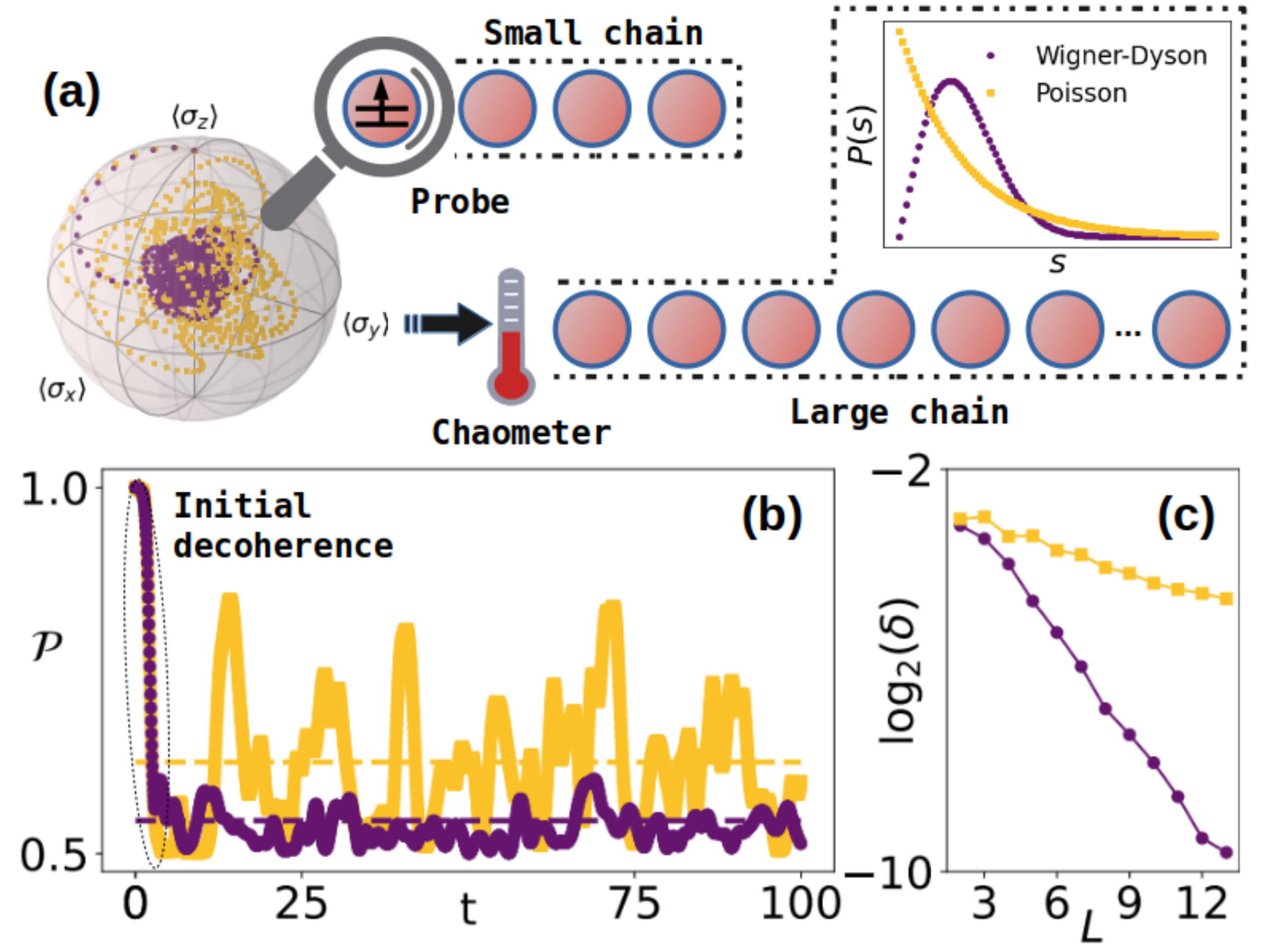}
\begin{footnotesize}
\caption{\nico{\nicor{Panel (a) Time evolution of the probe in the Bloch sphere}, considering a small chain of $L=6$ spins and both integrable (\nicor{$h_z=0.0$}, orange squares) and chaotic ($h_z=0.5$, violet circles) regimes. The initial state is a pure random state for each spin. The rest of the parameters are set as $h_x=1$, $J_k=1 \, \forall \, k=\{1,2...,L-1\}$ and \nicor{$T=100$} (in units of $J^{-1}$). Panel (b): Purity of the probe for the same set of parameters as (a). The temporal average is shown as a dashed line. \nicon{Panel (c): \nicof{Time} fluctuations of the purity (averaged over 50 different initial states) as a function of system size in both regimes. \nicof{Fluctuations are defined as $\delta(P)=\sqrt{\langle P(t)^2 \rangle - \langle P(t)\rangle ^2}$, where the interval $t \in [50,100]$ (in units of $J^{-1}$) was considered \cite{Note1}.}}}} 
\label{intro}
\end{footnotesize}
\end{center}
\end{figure}

\nico{In \nicon{panel (a) of}  Fig. \ref{intro} we summarize the main idea of our work. We are interested on how a single spin acting as a probe of a small chain behaves in the typical regimes where \nicof{the same spin chain but much larger} is known to be either integrable or chaotic.  With this purpose, \nicof{we solve} the Schrödinger equation for the whole \nicor{small} system $\rho(t)$ and then trace over the environmental degrees of freedom, focusing on the purity of the reduced density matrix $\tilde{\rho}(t)$ of the first spin of the chain ($\mathcal{P}(t)=\Tr[\tilde{\rho}^{2}(t)]$).} \nicon{Since the purity of the probe is fully determined through its Bloch vector $\vec{r}=(r_x,r_y,r_z)$, where $r_i(t)=\Tr \left(\sigma_i \tilde{\rho}(t) \right) \forall i \in \set{x,y,z}$ (i.e.
$\mathcal{P}(t)=1/2(1+|\vec{r}(t)|^2)$), its long-time dynamics is strictly related to the degree of equilibration of the whole set of local observables $A=\set{\mathrm{I},\sigma_x,\sigma_y,\sigma_z}$. This subsystem equilibration should be understand as $\lim\limits_{t \to \infty} \Tr (\tilde{\rho}(t)\widehat{O})=\Tr (\tilde{\rho}_{\infty} \widehat{O} ) \, \forall \, \widehat{O} \in A$, where $\tilde{\rho}_{\infty}$ is the equilibrium state of the probe \cite{linden2009quantum,short2012quantum}. \nicof{If the spin chain is large enough, under the assumption of infinite temperature}, we have $\tilde{\rho}_\infty=\frac{\mathrm{I}}{2}$, which implies $\Tr \left(\sigma_i \tilde{\rho}_\infty \right)=0 \, \forall i \in \set{x,y,z}$ and thus $\mathcal{P}_\infty=1/2$.}

\renewcommand{\figurename}{Figure} 

From \nicon{panel (b) of} Fig. \ref{intro} we can \nico{qualitatively} see that while in the chaotic regime the \nicor{long-time dynamics} washes out the purity of the system, leading to a state of almost maximum uncertainty, this is not the case for the integrable regime, where \nicor{at long times} it oscillates periodically around a mean value much greater than $1/2$. \nicon{It is clear that fluctuations are much smaller in the chaotic regime, despite the spin chain analyzed in Fig. \ref{intro} is quite short ($L=6$). Also, while fluctuations strongly decay with system size in this regime, they do not in the integrable case, as \nicof{it is} shown in panel (c) \nicof{of Fig. 1}.} \nicor{With respect to the short-time decay, associated with decoherence, it is similar in both regimes \nicof{\cite{ermann2006decoherence}}. For this reason, we will focus on the long-time regime, where \nicon{some degree of} equilibration takes place even in extremely short chains, as we shall see.}

\begin{figure}[!htb]
\begin{center}
\includegraphics[width=85mm]{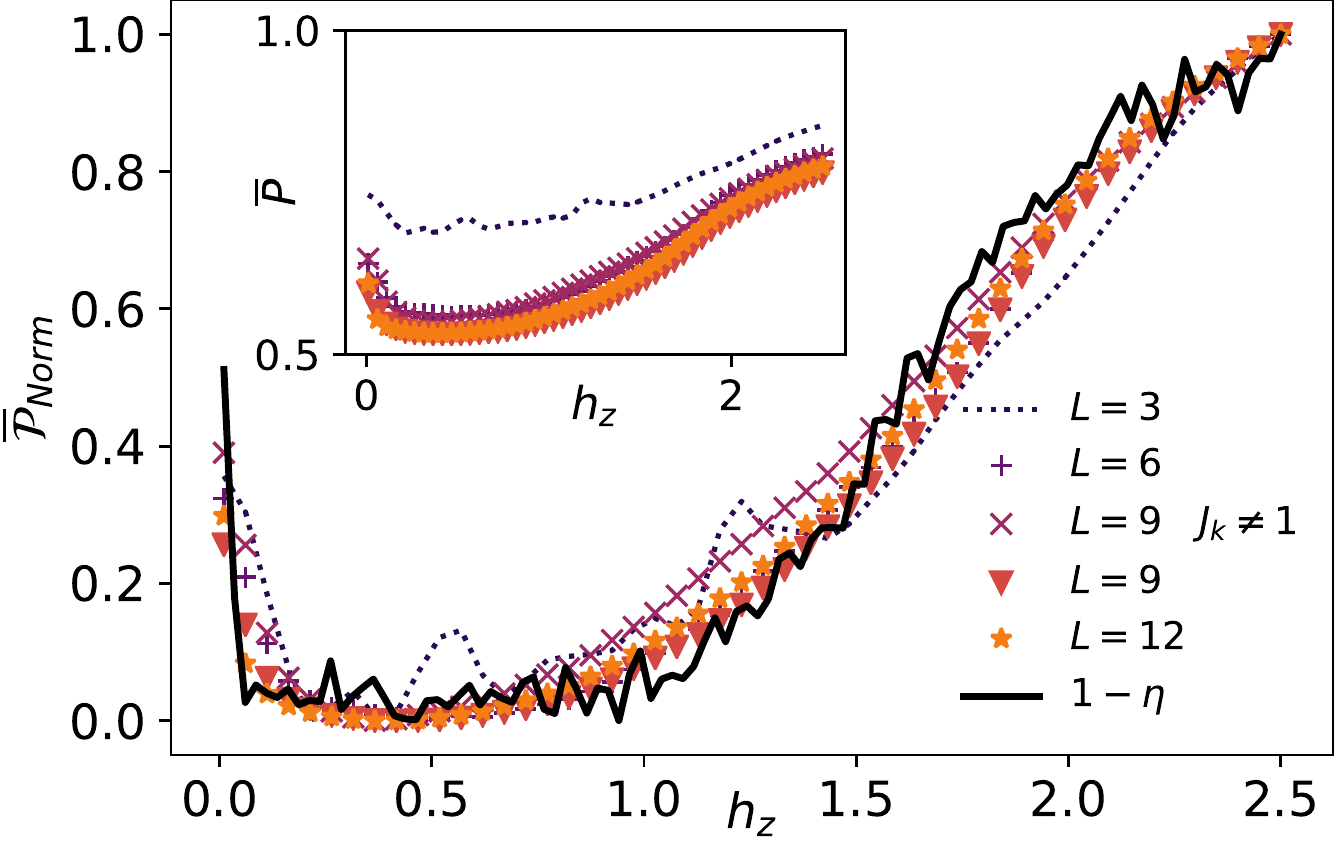}
\begin{footnotesize}
\caption{\textit{Main plot:} $\overline{\mathcal{P}}_{Norm}$ for the probe considering different sizes of the environment together with the chaos parameter $1-\eta$, both as a function of the magnetic field $h_z$. For computing $\overline{\mathcal{P}}_{Norm}$, 50 different realizations over random initial states were considered. For the calculation of $1-\eta$, a chain composed of $L=14$ spins ($D=16384$) was selected and only the odd subspace was taken into account ($D^{odd} \approx 8192$). Parameters are set as $T=50$, $h_x=1$ and $J_k=1 \, \forall \, k=\{1,2...,L-1\}$ with the exception of the violet crossed curve where $J_k \in [0.5,1.5] \, \forall \, k=\{1,2...,L-1\}$. \nicor{The plot begins at $h_z=0.01$}. \textit{Inset plot:} Same as the main plot but without normalizing the averaged purity (i.e. $\overline{\mathcal{P}}$).}
\label{chaos}
\end{footnotesize}
\end{center}
\end{figure}

Having this qualitative picture in mind, we now intend to measure \nicon{the degree of equilibration} in a more quantitative way. To do so, we will focus again on the purity degradation of our reduced spin system $\tilde{\rho}(t)$, by defining an averaged purity as
    $\overline{\mathcal{P}}=\dfrac{1}{N}\sum_{i=1}^{N}\left(\dfrac{1}{T}\int_{0}^{T}\Tr[\tilde{\rho}^{2}_{i}(t)]dt \right)$,
where we first make a temporal average over the purity of a particular $\tilde{\rho}_{i}(t)$, defined by a given random initial state, and then we repeat this procedure for $N$ different initial random states, to finally perform a global average over all realizations. \nico{Let us remark that since we are interested in studying the transition to chaos as a function of a certain parameter, to compare the averaged quantity $\overline{\mathcal{P}}$ with the chaos measure introduced in Eq. (\ref{eta}), we define a normalized averaged purity as}
\begin{equation}
    \overline{\mathcal{P}}_{Norm}=\dfrac{\overline{\mathcal{P}}-\min{(\overline{\mathcal{P}}})}{\max{(\overline{\mathcal{P}})}-\min{(\overline{\mathcal{P}})}
} \qquad \left( 0 \leq \overline{\mathcal{P}}_{Norm} \leq 1 \right),  
\label{purity}
\end{equation}
\nicor{where $\min{(\overline{\mathcal{P}})}$ and $\max{(\overline{\mathcal{P}})}$ are the minimal and maximal value obtained when sweeping over the parameter range.} With this definition, we have now all the necessary ingredients to \nicor{pose} the following question: how does the purity degradation of the reduced system behaves as a function of the degree of chaos present in the rest of the chain? To address this issue, in Fig. \ref{chaos} we plot the \nicon{spectral} chaos indicator $\eta$ for a large chain composed of $L=14$ spins ($D=16384$) together with the averaged purity $\overline{\mathcal{P}}_{Norm}$ of the reduced system for different sizes of the total spin chain, both as a function of the magnetic field $h_z$. 

\renewcommand{\figurename}{Figure} 

Interestingly, the behavior of the averaged purity of the probe is quite similar regardless of the length of the environment. In fact, there is a well distinguished area in all the curves where the purity degradation is maximal. By comparing with the curve given by $1-\eta$, we can see that this region coincides almost perfectly with the region where chaos reigns, i.e. $(1-\eta) \rightarrow 0$. Quite remarkably, this is true even when the \nicon{system} is extremely short ($D=8$), where we can observe a precise correspondence with the exception of a small deviation near $h_z \sim 0.5$. This deviation can be smoothed by either taking more realizations over different initial states or slightly increasing the size of the environment by one spin. 

Various implications emerge from the analysis of Fig. \ref{chaos}. In first place, by using one spin as a probe and studying its purity dynamics, we were able not only to sense the chaotic behaviour present in \nicon{the full system}, but also to reconstruct the whole integrable to chaos transition with a great degree of correspondence in comparison to other standard indicators of chaos. However, while the usual methods require a full diagonalization and classification of eigenenergies according to their symmetries within huge dimensional subspaces \cite{Note1}, we have obtained the same results without requiring the above and even in \nicon{much smaller subspaces}. Moreover, the average over different realizations of the purity proved to be robust not only to the size of the environment, but also to whether we consider equal couplings or even a random set of $J_k$ modelling some hypothetical experimental error (see violet crossed curve in Fig. \ref{chaos}). \nicon{Our result evidences that when a small fraction from a large chaotic system is selected, some trace of the universal nature of the latter survives.} 

Keeping in mind the results presented so far, let us now examine the following hypothetical situation: consider an experimental scenario where a \nico{given} spin chain \nicor{is well-isolated from the external environment and} where some particular spin of this chain can be externally controlled. For instance, consider a time-dependent Hamiltonian 
\begin{equation}
H=\sum_{k=1}^{L} \left(h_x \hat{\sigma}_{k}^{x}+h_z\hat{\sigma}_{k}^{z} \right) -   \sum_{k=1}^{L-1}J_{k}\hat{\sigma}_{k}^{z}\hat{\sigma}_{k+1}^{z}+\lambda(t)\hat{\sigma}_{1}^{z},    
\label{hamilt2}
\end{equation}
where $\lambda(t)$ is a control field that can be experimentally tuned. Thus,  you may want to implement some particular protocol over the spin you are able to control. For example, consider a population transfer protocol, where the first spin of the chain has to be addressed from the initial state $\ket{\psi(0)}=\ket{0}$ to the final target state $\ket{\psi_{targ}}=\ket{1}$. \nico{\nicon{Or maybe you are} interested in generating a maximally entangled state between the first two spins of the chain, i.e.  $\ket{\psi_{targ}}=\frac{1}{\sqrt{2}}\left( \ket{00} +\ket{11} \right)$}. To do so, the time-dependent control field $\lambda (t)$ must be optimized to maximize the fidelity $\mathcal{F}=\left |\braket{\psi(T)}{\psi_{targ}} \right |^{2}$ at a final evolution time $T$. In light of the results we presented before, you may be wondering the following question: is the maximum degree of control achievable subordinated to the degree of chaos present within the non-controlled \textit{environmental spins}?  

\begin{figure}[!htb]
\begin{center}
\includegraphics[width=85mm]{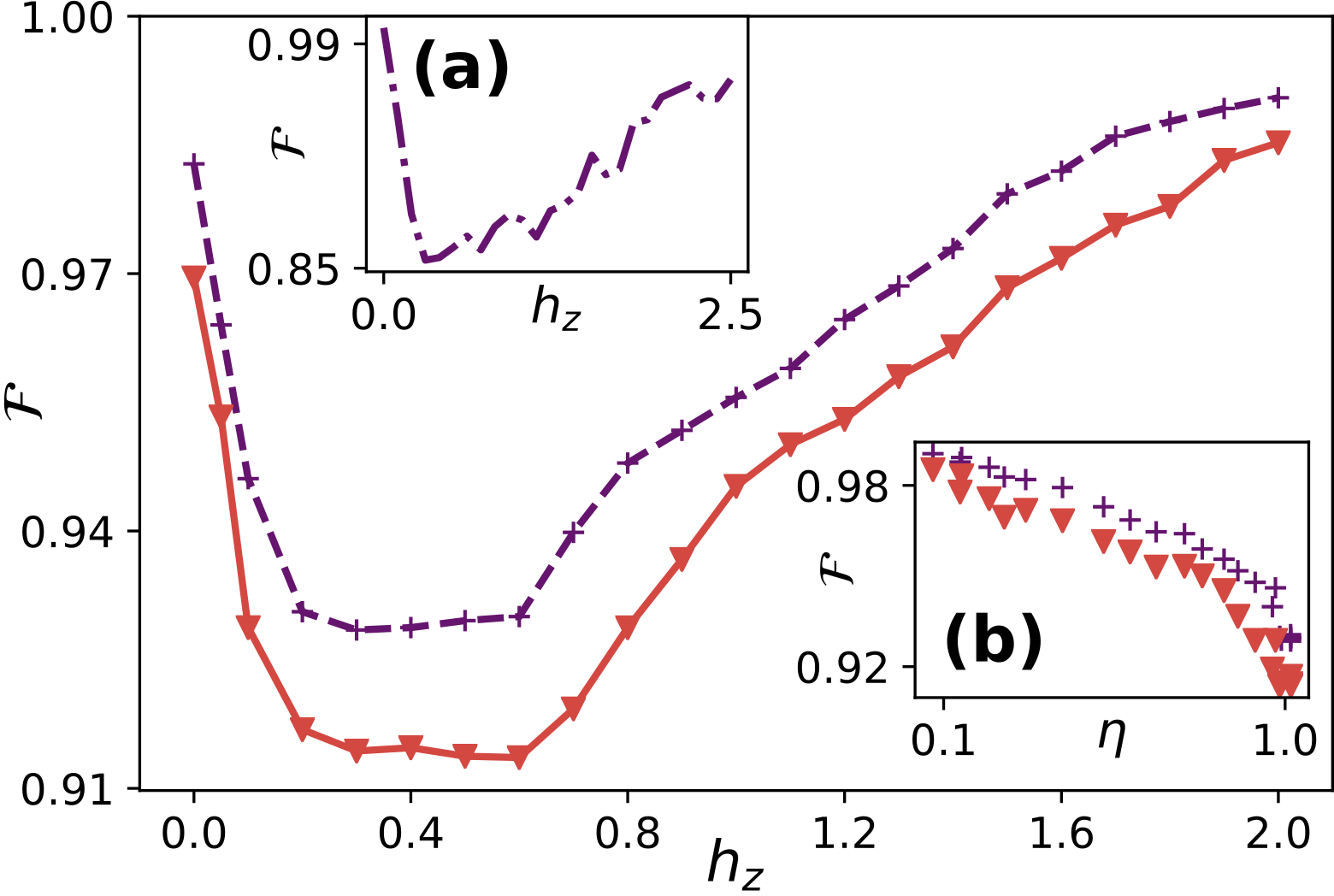}
\begin{footnotesize}
\caption{\textit{Main plot:} Optimal fidelities for a population transfer protocol as a function of $h_z$. The dashed curve is for $L=6$ spins and the solid for $L=9$. Interaction parameters are set as $T=20$, $h_x=1$ and $J_k \in [0.5,1.5] \, \forall \, k=\{1,2...,L-1\}$. The initial state is $\ket{0}$ for the first spin and random for the rest of the system (see Eq. (\ref{initial_state})). Only one realization was considered. \nico{\textit{Inset (a):} Optimal fidelities for an entangling protocol between the first two spins of the chain. Parameters are set as $L=6$, $T=20$, $h_x=1$ and $J_k =1 \, \forall \, k=\{1,2...,L-1\}$. The initial state is random for each spin and only one realization was considered.} \textit{\nico{Inset (b)}:} Optimal fidelities of the main plot as a function of the chaos parameter $\eta$.} 
\label{control}
\end{footnotesize}
\end{center}
\end{figure}

In order to answer \nicof{this} question, we \nicof{consider} the control function $\lambda(t)$ as a vector of control variables $\lambda(t) \rightarrow \{ \lambda_{l} \} \equiv \Vec{\lambda}$, i.e. a field with constant amplitude $\lambda_{l}$ for each time step. \nico{By dividing the evolution time $T$ into $n_{ts}$ equidistant time steps  $(l=1,2,..., n_{ts})$, the optimization was performed exploring several random initial seeds and resorting to standard optimization tools \cite{bib:scipy, johansson2012qutip}. In Fig. \ref{control} we plot the optimal fidelities achieved for both the population transfer and entangling protocols, as a function of $h_z$ and for different lengths for the total spin chain.} 
\renewcommand{\figurename}{Figure} 

Interestingly, we can conclude from Fig. \ref{control} that the optimal fidelities achieved for \nicon{these simple but paradigmatic protocols} \nicof{are} very sensitive to the degree of chaos that is present within the rest of the spin chain. In fact, from the main plot \nico{and from inset (a)} we can see that the optimal fidelities behave quite similarly to the chaos parameter $1-\eta$, as a function of the magnetic field $h_z$ (see Fig. \ref{chaos}). Accordingly, in the \nico{inset (b)} of Fig. \ref{control} we plot the optimal fidelities obtained in the main plot but now as a function of the degree of chaos associated to the specific strength of the magnetic field $h_z$ (see again Fig. \ref{chaos}). By doing this, it is clear that the more chaos, the worse control. This last statement clearly relates to what we have been discussing before, in the sense that a greater degree of chaos is also associated with a stronger \nicor{equilibration}. Therefore, this means that the non-controlled system is acting as an \textit{effective environment} for the \nico{spins that are} being actively controlled and we argue that even in the case where this effective environment is small, its dynamics should be carefully tuned in order to minimize \nicon{equilibration} and thus improve the degree of control over the reduced system that is being addressed.

\textit{Concluding remarks.} The goal of this work was to \nicor{study} the interplay between equilibration, quantum chaos and control in the limit \nicon{of a small isolated many-body quantum system}. \nicon{In this context, by monitoring the long-time dynamics of a spin connected to a generic spin chain, we found that its purity degradation can be used as a probe to sense the chaotic behaviour of the chain \nicof{under} the limit of infinite temperature. By showing that a greater degree of equilibration is associated with a more chaotic region, we were able to reconstruct the whole integrable to chaos transition even in the case where the full system was merely composed of three spins.} \nico{This was done without any consideration \nicof{of} the conserved symmetries of the system, which is \nicof{another} important advantage with respect to previous methods considered in the literature.} \nicon{\nicof{The fact of finding robust vestiges of quantum chaos in such small quantum systems is of fundamental interest but also has practical implications in quantum control experiments}}. By considering \nicon{simple but paradigmatic protocols} over a spin subject to a control field that can be experimentally tuned, we have shown that the best control achievable is a function of the degree of chaos present \nicon{within the full system of which it is a part}. Consequently, in realistic experiments where a control task is sought over a reduced part of a system that is not necessarily large but that nevertheless presents signatures of quantum chaos, the interaction parameters must be carefully adjusted to avoid the chaotic regime and thus achieve a better performance of the control. 

\begin{acknowledgements}
We acknowledge R. A. Jalabert for his insights about the manuscript. The work was partially supported by CONICET (PIP 112201 50100493CO), UBACyT (20020130100406BA), ANPCyT (PICT-2016-1056), and National Science Foundation (Grant No. PHY-1630114).
\end{acknowledgements}

\bibliography{main.bib}

\widetext
\clearpage

\begin{center}
\textbf{\large Supplemental Material}
\end{center}
\setcounter{equation}{0}
\setcounter{figure}{0}
\setcounter{table}{0}
\setcounter{page}{1}
\makeatletter
\renewcommand{\theequation}{S\arabic{equation}}
\renewcommand{\thefigure}{S\arabic{figure}}
\renewcommand{\bibnumfmt}[1]{[S#1]}
\renewcommand{\citenumfont}[1]{S#1}

\nico{\section{Other physical systems}
\subsection{Ising with tilted magnetic field}
\label{sec_ising_tilt}
This model consists of an Ising spin chain with a tilted magnetic field. It undergoes a quantum chaos transition for intermediate angles between the purely longitudinal and the transverse case \cite{karthik2007entanglement}. The Hamiltonian of the system with open boundary conditions is given by

\begin{equation}
H (\theta) =B \sum_{k=1}^{L} \left(\sin{\theta} \hat{S}_{k}^{x}+\cos{\theta}\hat{S}_{k}^{z} \right) + J \sum_{k=1}^{L-1}\hat{S}_{k}^{z}\hat{S}_{k+1}^{z}.
\label{tilted}
\end{equation}

Since parity $\hat{\Pi}$ is also a conserved symmetry in this model, even and odd subspaces must be treated separately to study the usual chaos indicators as $\eta$. \nicor{For the purity analysis of the probe, we proceed exactly the same as in the main text but with a single variation: we assume that the probe does not evolve at all in the absence of coupling to the bath, i.e. we neglect its intrinsic Hamiltonian and just consider its coupling to the rest of the chain as its full Hamiltonian. This detail is important to avoid an asymmetry with respect to the ensemble of initial states under consideration and allows us to sense more accurately the chaos within the chain, as we shall see.} In Fig. \ref{ising_tilt} we show the results of the chaos indicator $\eta$ for a large chain composed of $L=14$ spins ($D=16384$) together with the averaged purity $\overline{\mathcal{P}}_{Norm}$ of the reduced system \nicor{(with and without considering the intrinsic Hamiltonian of the probe)} for different sizes of the total spin chain, both as a function of the magnetic field angle $\theta$.  

\renewcommand{\figurename}{Figure} 
\begin{figure}[!htb]
\begin{center}
\includegraphics{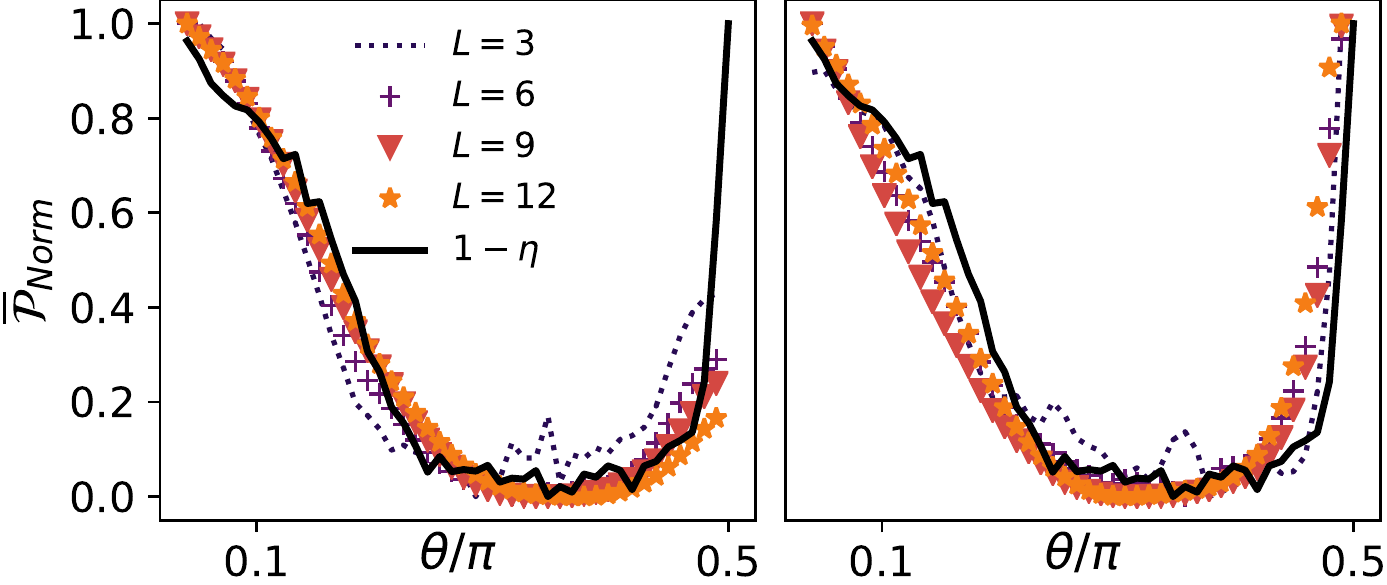}
\begin{footnotesize}
\caption{\nico{Normalized averaged purity $\overline{\mathcal{P}}_{Norm}$ for the reduced system considering different sizes of the environment together with the chaos parameter $1-\eta$, both as a function of the magnetic field angle $\theta$. For the computation of $\overline{\mathcal{P}}_{Norm}$, 50 different realizations over random initial states were considered. For the calculation of $1-\eta$, a chain composed of $L=14$ spins ($D=16384$) was selected and only the odd subspace was taken into account ($D^{odd} \approx 8192$). Parameters are set as $T=50$, $B=2$ and $J=2$.} \nicor{While in the left panel we include the Hamiltonian of the probe, in the right panel we do not.}} 
\label{ising_tilt}
\end{footnotesize}
\end{center}
\end{figure}

\nicor{From the right panel of} Fig. \ref{ising_tilt}, we can see that even in the case of $L=3$, by studying the purity dynamics of one spin of the chain, we can perfectly reproduce the integrable to chaos transition. Unlike the computation of $\eta$, no symmetry considerations were taken into account for the analysis of the purity. \nicor{On the other hand, from comparing both panels it is evident that the fact of neglecting the intrinsic Hamiltonian of the probe (right panel) gave us a better estimation than in the case of considering it (left panel), mostly near the limit of $\theta \eqsim \frac{\pi}{2}$. This can be understood as follows: Let us consider the Hamiltonian of Eq. (\ref{tilted}) including the intrinsic Hamiltonian of the probe. It is easy to see that if $\theta=0$, Eq. (\ref{tilted}) is purely in the $\hat{z}$ direction and thus the $2^{L}$ eigenstates of the total Hamiltonian are the ones of the computational basis. If we consider these eigenstates and trace over the environmental degrees of freedom, the \textit{'reduced eigenstate'} of the probe is either $\ket{0}$ or $\ket{1}$, which are both pure. Therefore, if the random pure initial state of the probe falls near $\ket{0}$ or $\ket{1}$, the probe will hardly evolve since it is near an eigenstate of the Hamiltonian. On the contrary, let us consider now the opposite limit of $\theta=\pi/2$, where the system is also integrable. By diagonalizing the entire Hamiltonian and then tracing over the environmental degrees of freedom, the \textit{'reduced eigenstates'} of the probe are in general not pure. As a consequence, the initial state of the probe, which is a random pure state, is never going to fall near a \textit{'reduced eigenstate'} (since it is mixed), and so all possible initial states will strongly evolve in time. This asymmetry with respect to the initial state of the probe in the two extreme cases of integrability ($\theta=0$ and $\theta=\pi/2$), can be trivially solved by neglecting the intrinsic Hamiltonian of the probe in the model. By doing this, the \textit{'reduced eigenstate'} of the probe is always pure $\forall \, \theta$ (i.e. $\ket{0}$ and $\ket{1}$) and the asymmetry is fixed, as it is clearly evidenced in the right panel of Fig. \ref{ising_tilt}. However, as has been noted before, both panels are quite similar, with the exception of $\theta \eqsim \pi/2$.}

\subsection{Heisenberg with random magnetic field}
This system consists of a spin chain with nearest neighbour interactions (NN), coupled to a random magnetic field at each site in the $\hat{z}$ direction \cite{vznidarivc2008many}. The Hamiltonian for this model with open boundary conditions is 

\begin{equation}
H=\sum_{k=1}^{L-1}\left(\hat{S}_{k}^{x}\hat{S}_{k+1}^{x} + \hat{S}_{k}^{y}\hat{S}_{k+1}^{y} + \hat{S}_{k}^{z}\hat{S}_{k+1}^{z} \right) + \sum_{k=1}^{L} h_k^z \hat{S}_{k}^{z},     
\label{heisenberg}
\end{equation}
where $\{h_k^z\}$ is a set of random variables at each site, uniformly distributed in the interval $[-h,h]$. In this model, the $\hat{z}$ component of spin $\hat{S}^z=\sum_{k=1}^L \hat{S}_k^z$ is a conserved quantity. This conservation allows a separation of the total spanned space into smaller subspaces of dimension $\hat{\mathcal{S}}_N$, where $N$ is a fixed number of spins up or down. The dimension of each subspace is given by

\begin{equation}
    D_N= \dim\left(\hat{\mathcal{S}}_N\right)= \begin{pmatrix}
L\\
N 
\end{pmatrix}
=\frac{L!}{N!(L-N)!}.
\end{equation}

Taking into consideration the symmetry mentioned, it can be shown that this model presents a Poissonian distribution for $h=0$. However, as the strength of the random magnetic field increases, the disorder is higher and the system is driven to the chaotic regime, reaching a Wigner-Dyson distribution near $h \simeq 0.5$. At last, if the disorder is too strong, there is a  many-body localization (MBL) transition \cite{avishai2002level,santos2004integrability,pal2010many,de2013ergodicity,luitz2015many}. In Fig. \ref{heisenberg} we show the results for the averaged chaos indicator $\eta$ for a large chain composed of $L=13$ spins, after performing 50 different realizations due to the stochastic nature of the random set of $\{h_k^z\}$. In this simulation, just the subspace with $N=5$ was considered ($D_5=1287$). As usual, we also plot the averaged purity $\overline{\mathcal{P}}_{Norm}$ of the reduced system for different sizes of the total spin chain, without any consideration of the symmetries present in the system. Both quantities are plotted as a function of the strength of the magnetic field $h$. 
\renewcommand{\figurename}{Figure} 
\begin{figure}[!htb]
\begin{center}
\includegraphics{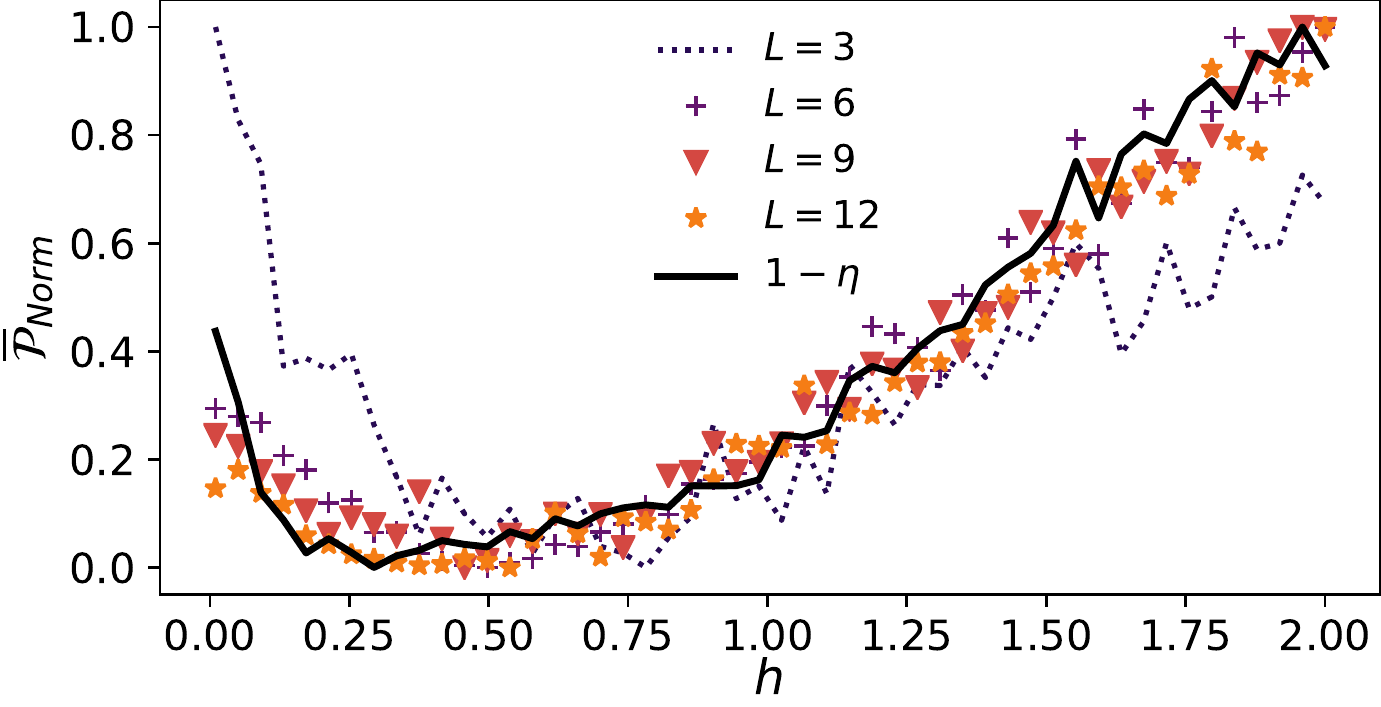}
\begin{footnotesize}
\caption{\nico{Normalized averaged purity $\overline{\mathcal{P}}_{Norm}$ for the reduced system considering different sizes of the environment together with the chaos parameter $1-\eta$, both as a function of the magnetic field  $h$. For the computation of $\overline{\mathcal{P}}_{Norm}$, $200$ different realizations over random initial states and random sets of $\{h_k^z\}$ were considered. For the calculation of $1-\eta$, $50$ different realizations over random sets of $\{h_k^z\}$ were considered, a chain composed of $L=13$ spins was selected and only the subspace with $N=5$ was taken into account ($D_5 \approx 1287$). Parameters are set as $T=50$.}} 
\label{heisenberg}
\end{footnotesize}
\end{center}
\end{figure}
Quite remarkably, despite the system presents completely different symmetries and interactions in comparison to the other models that were studied in this work, the purity dynamics of a simple probe still senses the chaotic behaviour of the chain very accurately. Since in this model we have two different sources of randomness (i.e. the initial state and the magnetic field at each site are both random), is natural to observe some more noise especially for extremely short spin chains as the case of $L=3$. This noise can be smoothed out by taking more realizations over more different initial states and sets of $\{h_k^z\}$, but we can observe that even for $200$ realizations the transition can be very well detected analyzing extremely short spin chains. Interestingly, this system presents a MBL transition for strong disorder, but the interplay between the proposed measure and MBL will be addressed in a future work.   

\newpage
\subsection{Perturbed XXZ model}
The last spin chain to be analyzed in this work consists on an anisotropic spin chain with nearest-neighbor (NN) interactions and a perturbation consisting in next-nearest-neighbor (NNN) interactions. The total Hamiltonian of the system with open boundary conditions is described by

\begin{equation}
H(\lambda)=H_0 + \lambda H_1,
\end{equation}
where the parameter $\lambda$ tunes the strength of the perturbation and each term of the total Hamiltonian is
\begin{equation}
    \begin{split}
        & H_0= \sum_{k=1}^{L-1}\left(\hat{S}_{k}^{x}\hat{S}_{k+1}^{x} + \hat{S}_{k}^{y}\hat{S}_{k+1}^{y} +\mu \hat{S}_{k}^{z}\hat{S}_{k+1}^{z} \right) \\
        & H_1= \sum_{k=1}^{L-2}\left(\hat{S}_{k}^{x}\hat{S}_{k+2}^{x} + \hat{S}_{k}^{y}\hat{S}_{k+2}^{y} + \mu \hat{S}_{k}^{z}\hat{S}_{k+2}^{z} \right).
    \end{split}
\end{equation}

This system presents several symmetries that must be taken into account to observe the quantum chaos transition in a standard way \cite{santos2012onset}. This transition occurs when the perturbation term containing NNN interactions becomes comparable with the NN coupling term. If the chain is isotropic ($\mu=1$) the total spin $\hat{S}^2$ is conserved, so we must consider a case with $\mu \neq 1$ to see the transition. At the same time, this system not only presents a conservation in the $\hat{z}$ component of spin $\hat{S}^z=\sum_{k=1}^L \hat{S}_k^z$ but also parity is a conserved quantity. Therefore, we must consider subspaces with a fixed number of spin up or down but also separate them according to their parity. With such considerations, in Fig. \ref{xxZ} we show our analysis comparing the spectral indicator $\eta$ with the purity dynamics of our probe, for different sizes of the total spin chain, both as a function of the perturbation strength $\lambda$. While for computing $\eta$ we have considered a spin chain of length $L=15$ and restricted ourselves to the even subspace of $N=5$ spins up, for the analysis concerning the purity dynamics of our probe, no consideration regarding symmetry was made.  
\renewcommand{\figurename}{Figure} 
\begin{figure}[!htb]
\begin{center}
\includegraphics{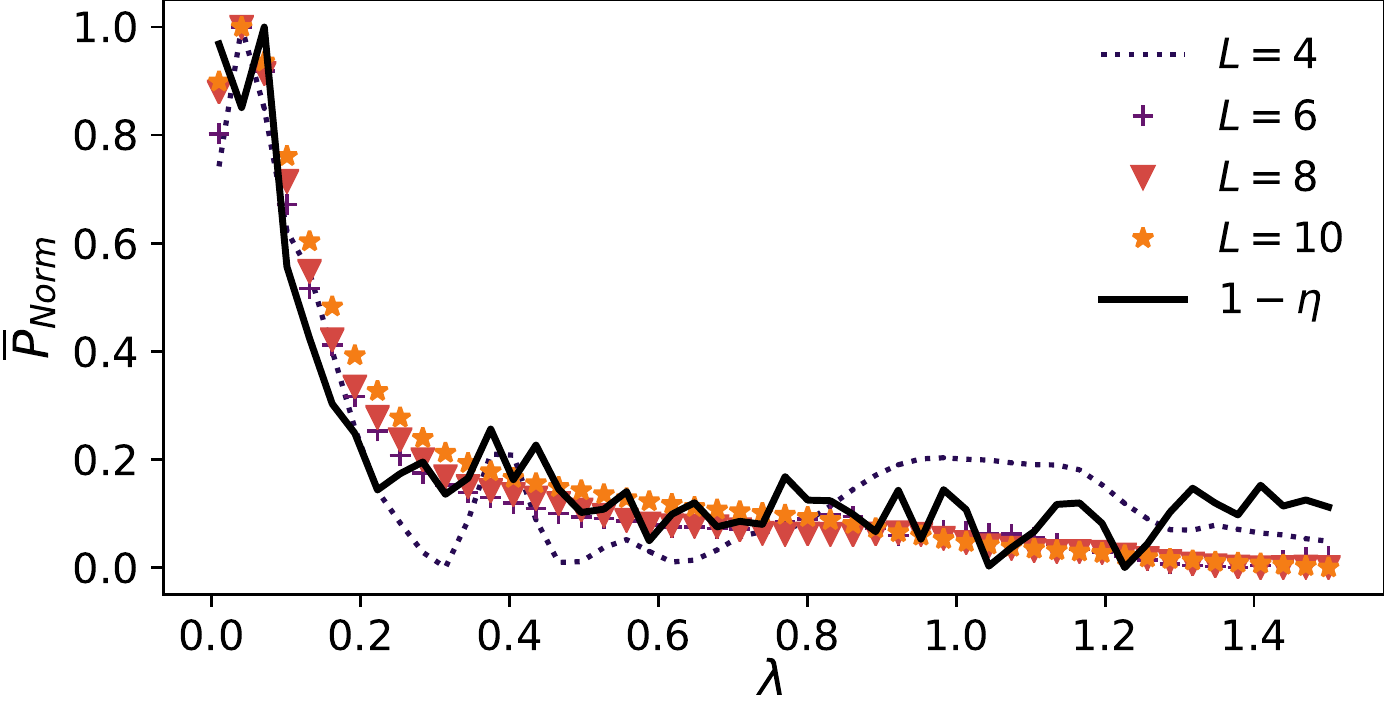}
\begin{footnotesize}
\caption{\nico{Normalized averaged purity $\overline{\mathcal{P}}_{Norm}$ for the reduced system considering different sizes of the environment together with the chaos parameter $1-\eta$, both as a function of the perturbation strength $\lambda$. For the computation of $\overline{\mathcal{P}}_{Norm}$, $50$ different realizations over random initial states were considered. For the calculation of $1-\eta$, a chain composed of $L=15$ spins was selected and only the subspace with $N=5$ was taken into account ($D_5 \approx 1500$). Parameters are set as $T=50$, $\mu=0.1$.}} 
\label{xxZ}
\end{footnotesize}
\end{center}
\end{figure}
Despite all the symmetries mentioned before, once again the purity degradation of the probe is enough to sense the chaotic behaviour in extremely short spin chains as the one of $L=4$ spins shown in the plot.  }

\clearpage

\section{Sensitivity of standard indicators to the dimension of Hilbert space}
\begin{figure}[!htb]
\begin{center}
\includegraphics{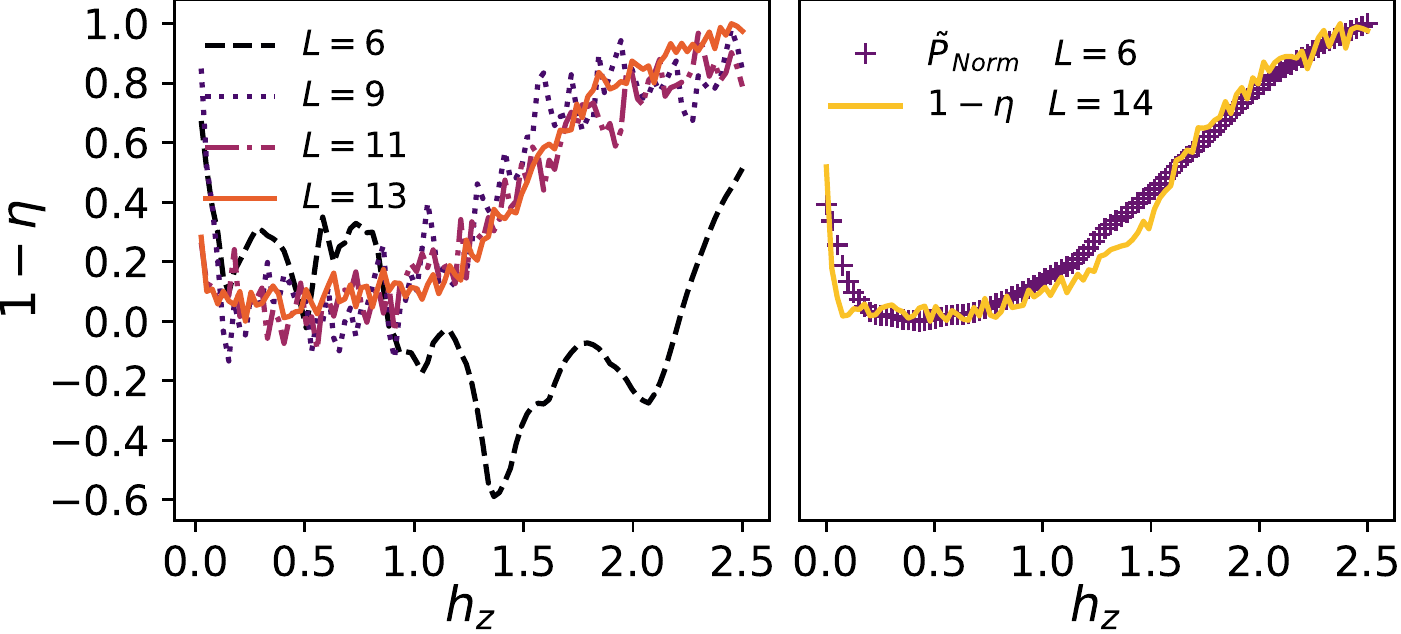}
\begin{footnotesize}
\caption{\textit{Left panel}: The indicator $1-\eta$ as a function of the magnetic field $h_z$ for different sizes of the total system. Only the odd subspace was taken into account for the calculation. \textit{Right panel}: Same indicator $1-\eta$ for a case of $L=14$ spins ($D=16384$) together with $\mathcal{\tilde{P}}_{Norm}$ for a case of $L=6$ spins ($D=64$), both as a function of the magnetic field $h_z$. In both panels, interaction parameters are set as $h_x=1$, $J_k=1 \, \forall \, k$, $T=50$. For the computation of $\mathcal{\tilde{P}}_{Norm}$, 50 realizations considering different random initial pure states were performed. }
\label{noise_eta}
\end{footnotesize}
\end{center}
\end{figure}

It is well-known that the usual chaos indicators are quite sensitive to the dimension of the Hilbert space considered. Indeed, laborious diagonalizations over large subspaces are required to have reasonable statistics of the energy levels. For instance, to calculate $1-\eta$ (see Fig. 2) we have resorted to a diagonalization within a dimension of $D=16384$ eigenstates and then classified them according to their symmetries. Despite the huge dimension used for the calculation, this measure has the disadvantage of still being noisy. As it is shown in the left panel of Fig. \ref{noise_eta} \nico{for the spin chain treated in the main text (is analogous for the other models)}, this noise is quite sensitive to the reduction of dimensionality. If we slightly decrease the size of the system on which we perform the  statistics, the fluctuations become larger and larger. However, this is not the case for $\mathcal{\tilde{P}}_{Norm}$, which is robust to the size of the system under consideration. As can be seen from Fig. \ref{noise_eta}, as we increase the size of the system for the calculation of $1-\eta$, the curve slowly tends to the one obtained by studying the purity dynamics of a system of much smaller dimensionality.

\renewcommand{\figurename}{Figure} 

\label{appA}

\newpage

\section{Average over different realizations vs a single realization}
\label{appB}

\begin{figure}[!htb]
\begin{center}
\includegraphics{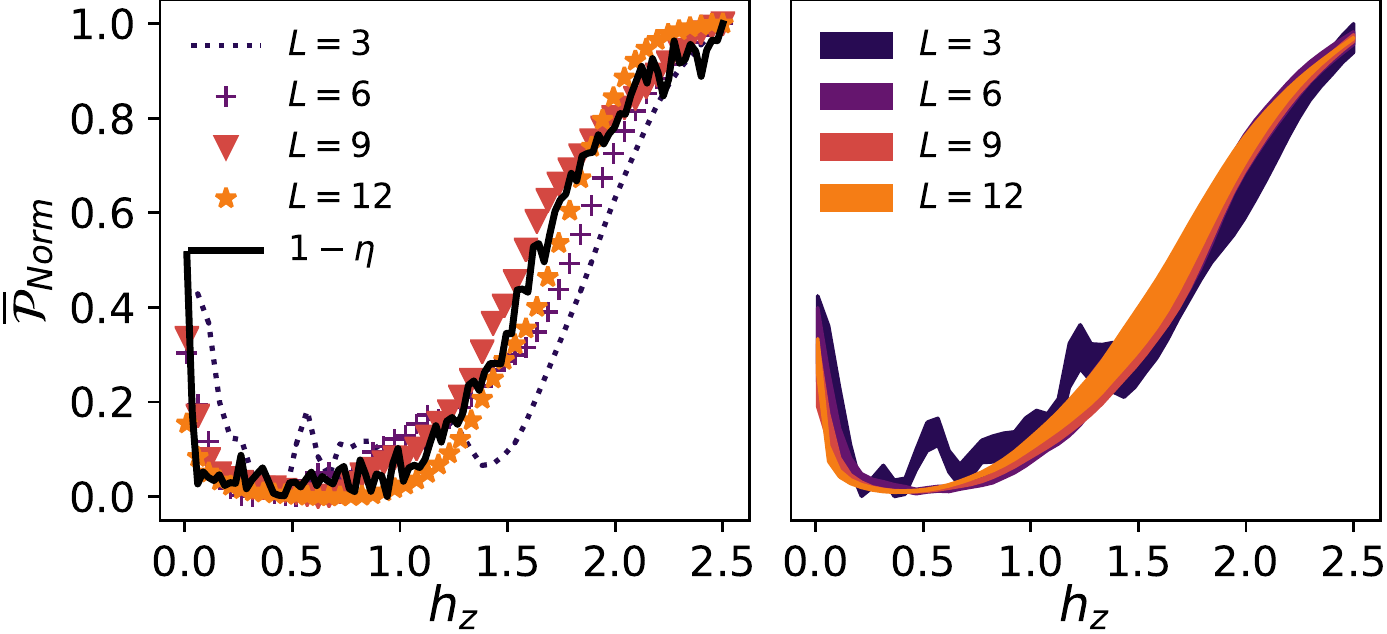}
\begin{footnotesize}
\caption{\textit{Left panel:} Normalized averaged purity $\overline{\mathcal{P}}_{Norm}$ for a single realization considering different sizes of the environment together with the chaos parameter $1-\eta$, both as a function of the magnetic field $h_z$. For computing $1-\eta$, a chain composed of $L=14$ spins ($D=16384$) was selected and only the odd subspace was taken into account ($D^{even} \approx 8192$). \textit{Right panel:} \nicor{Normalized averaged purity $\overline{\mathcal{P}}_{Norm}$}, as a function of the magnetic field $h_z$. The linewidth of each curve represents the variance considering 50 different realizations of random initial pure states. In both panels, interaction parameters are set as $h_x=1$, $J_k=1 \, \forall \, k=\{1,2...,L-1\}$ and $T=50$.}
\label{one_realization}
\end{footnotesize}
\end{center}
\end{figure}

Since the standard chaos measures rely on the spectral and statistical properties of the entire system, being independent of the initial state, we have defined the quantity $\mathcal{\tilde{P}}_{Norm}$ as an average over several random initial pure states such as to avoid privileging any particular initial condition \nicon{and thus sense the whole spectrum with almost the same probability}. Nevertheless, it is also important to analyze in general what is the behaviour of this quantity for a single realization and verify that the variance of the averaged quantity is small. This is precisely what is shown in Fig. \ref{one_realization} for different sizes of the total spin chain \nico{treated in the main text}. We can see that even a single realization is sensitive to the integrable to chaos transition and the variance of the averaged quantity is in general quite small. We remark that this is consistent with our result illustrating the implications in quantum control experiments (see Fig. 3), where only a single realization considering a fixed random pure initial state was considered. 

\renewcommand{\figurename}{Figure}

\newpage

\nicon{\section{Equilibration dynamics: Fluctuations and system size}
In the main text we have linked the notion of purity loss, equilibration and \nicof{quantum chaos at} the limit of infinite temperature and extremely small spin chains. Under this limit, depending on how small the quantum system effectively is and the interaction parameters that are set, a stronger or weaker degree of equilibration should be expected. This degree of equilibration is characterized by how far from the equilibrium state the quantum system is and how big fluctuations are at long times. \nicof{In this context, since the purity of a two-level system is determined through its Bloch vector} 

\begin{equation}
    \mathcal{P}(t)=\frac{1}{2}(1+|\vec{r}(t)|^2),
\end{equation}
where $r_i(t)=\Tr(\sigma_i \tilde{\rho}(t)) \, \forall i \in \set{x,y,z}$, the purity fluctuations are strongly correlated \nicof{to} the fluctuations of each component of $\vec{r}$. As well, these fluctuations can be associated with the non-diagonal elements of the expectation value of the observable $\vec{O}=\vec{\sigma} \otimes \mathrm{I}_B$, where $\mathrm{I}_B$ represents the identity acting over the whole \nicof{spin chain that} the probe is sensing, i.e. 

\begin{equation}
    \vec{r}(t) = \Tr (\vec{\sigma} \tilde{\rho}(t))= \Tr ( \vec{\sigma} \otimes \mathrm{I}_B \rho(t))= \sum_n |C_n|^2 \bra{n}\vec{\sigma} \otimes \mathrm{I}_B \ket{n}
    + \sum_{m\neq n} C_n C^*_m  \bra{m}\vec{\sigma} \otimes \mathrm{I}_B \ket{n}  e^{-i(E_n-E_m)t},
\end{equation}
where $\ket{n}$ are the eigenstates of the full system, $E_n$ its eigenenergies and $C_n=\bra{n}\psi(0)\rangle$. 
\nicof{Under} the limit of infinite temperature, the equilibrium state of the probe is $\tilde{\rho}_{\infty}=\frac{\mathrm{I}}{2}$, $|\vec{r}_\infty|=0$ and thus $\mathcal{P}_{\infty}=1/2$. 
The degree of equilibration to this state relies on how appreciable fluctuations are at long times and we have already shown in the main text that these fluctuations are much smaller in the chaotic regime, implying a greater degree of equilibration with respect to the integrable case (see Fig. 1). This is also exhibited in Fig. \ref{fluct_fig1}, where we can see that while fluctuations in the chaotic regime strongly decay \nicof{as we increase} L (left panel), they \nicof{vary smoothly} in the integrable case (right panel).    

\renewcommand{\figurename}{Figure}
\begin{figure}[!htb]
\begin{center}
\includegraphics{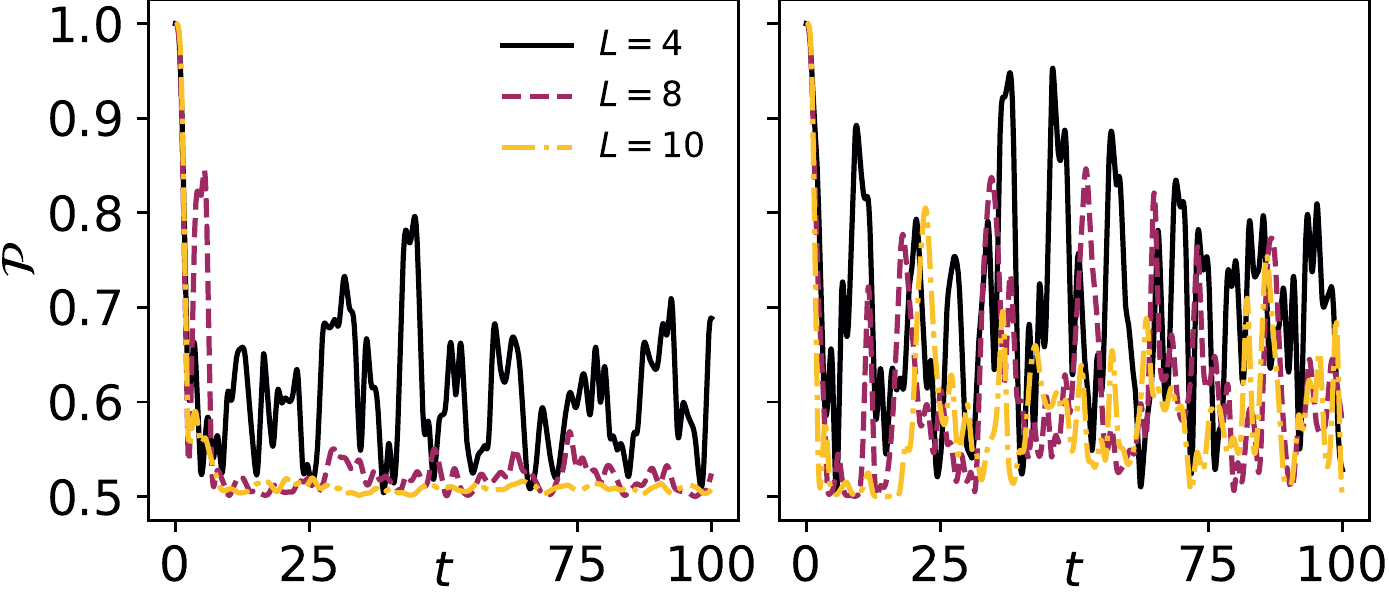}
\begin{footnotesize}
\caption{\nicon{Left panel: Purity as a function of time for three different system sizes. Parameters are set as $T=100$ (in units of $J^{-1}$), $h_z=0.5$, $h_x=1, J_k=1 \forall k$ (chaotic regime). Right panel: Same as the left panel but setting $h_z=0.0$ (integrable regime). In both panels, the same single realization over a pure random initial state was considered for each system size.}}
\label{fluct_fig1}
\end{footnotesize}
\end{center}
\end{figure}

To show the latter in a more rigorous way, we can define time fluctuations of a given quantity $X$ by $\delta(X)=\sqrt{\langle X(t)^2 \rangle - \langle X(t)\rangle ^2}$, where the  averages are taken over  appropriate time windows. With this definition, in Fig. \ref{fluct_fig} we plot the fluctuations of each component of the Bloch vector of the probe as a function of system size in both regimes. We can see that in the chaotic regime (left panel), fluctuations associated with the expectation value of each component of the observable $\vec{O}=\vec{\sigma}\otimes \mathrm{I_B}$ decay exponentially with system size. More explicitly, this scaling is related to the thermodynamic entropy, which under our assumption of infinite temperature is $S(E)=\ln (2^L)$ and thus $e^{-S(E)/2}=e^{-\ln{2^L}/2}=1/(2^{L/2})$. On the contrary, in the integrable regime (right panel), fluctuations also decay but much less dramatically than in the chaotic regime. This means that the degree of equilibration experienced by the probe in the integrable regime is much smaller, as was also pointed out in several parts of the main text.

\renewcommand{\figurename}{Figure}
\begin{figure}[!htb]
\begin{center}
\includegraphics{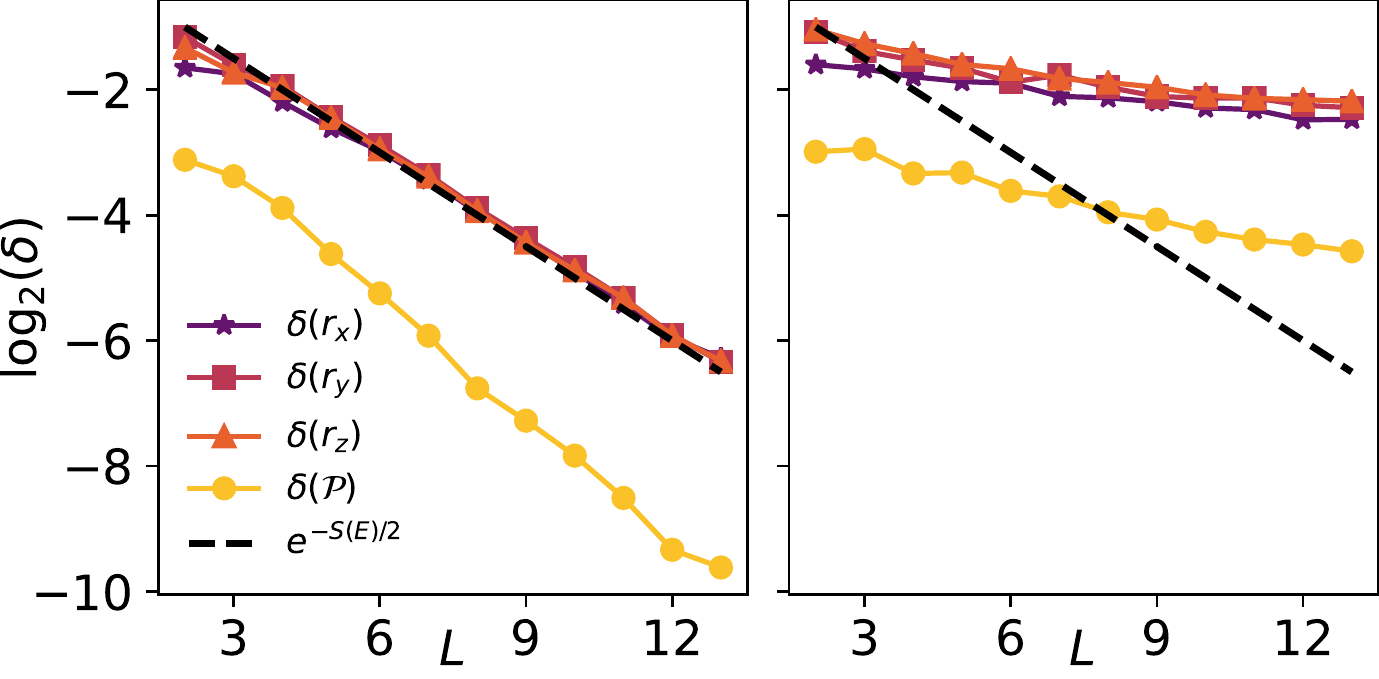}
\begin{footnotesize}
\caption{\nicon{Left panel: Fluctuations of the purity and of each component of the Bloch vector of the probe $r_i \, \forall i \in \set{x,y,z}$  between the time interval $t \in [50,100]$ (in units of $J^{-1}$). Parameters are set as $h_z=0.5$, $h_x=1, J_k=1 \forall k$ (chaotic regime). Right panel: Same as the left panel but setting $h_z=0.0$ (integrable regime). In both panels and for each system size, 50 different realizations over pure random initial states were considered (the mean over all realizations is what is shown).} 
}
\label{fluct_fig}
\end{footnotesize}
\end{center}
\end{figure}
}

\clearpage

\begin{figure}[!htb]
\begin{center}
\includegraphics{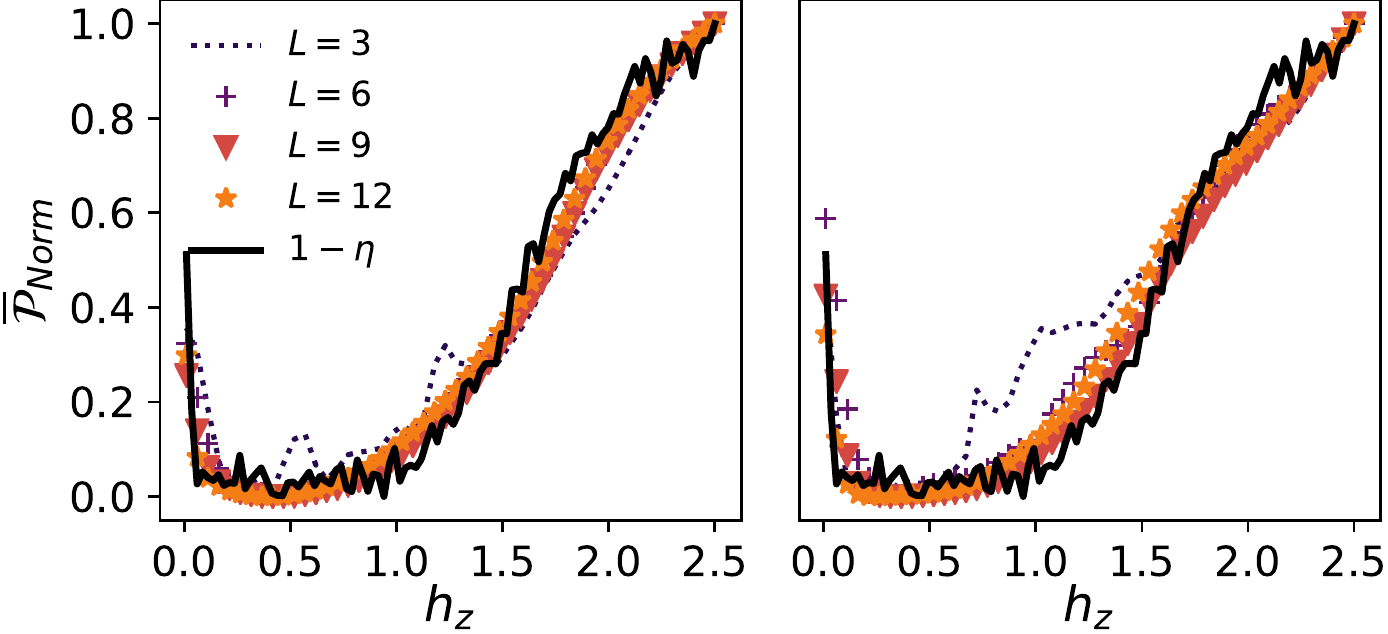}
\begin{footnotesize}
\caption{\nicon{Normalized averaged purity $\overline{\mathcal{P}}_{Norm}$ for the probe considering different sizes of the environment together with the chaos parameter $1-\eta$, both as a function of the magnetic field $h_z$. Interaction parameters are set the same as in Fig. 2 of the main text and 50 different realizations over random initial states were considered. \textit{Left panel}: Including the intrinsic Hamiltonian of the probe (same curves as in Fig. 2 of the main text). \textit{Right panel}: Neglecting the intrinsic Hamiltonian of the probe. }}
\label{other_setups}
\end{footnotesize}
\end{center}
\end{figure}

\nicon{\section{Different sensing setups: probe Hamiltonian and finite environmental temperatures}}
\nicon{Here we analyze different possibilities for sensing the chaotic behaviour of the Ising spin chain presented in the main text. As we have already discussed on Section \ref{sec_ising_tilt}, one possibility is to neglect the intrinsic Hamiltonian of the probe. In Fig. \ref{other_setups} we compare the results for the same setup considered in the main text (left panel) with a case where the intrinsic Hamiltonian of the probe is neglected (right panel). As can be observed, both situations lead to quite similar results even in the extreme case of $L=3$ spins, which evidences the robustness of the method proposed.} 

\renewcommand{\figurename}{Figure}

\begin{figure}[!htb]
\begin{center}
\includegraphics{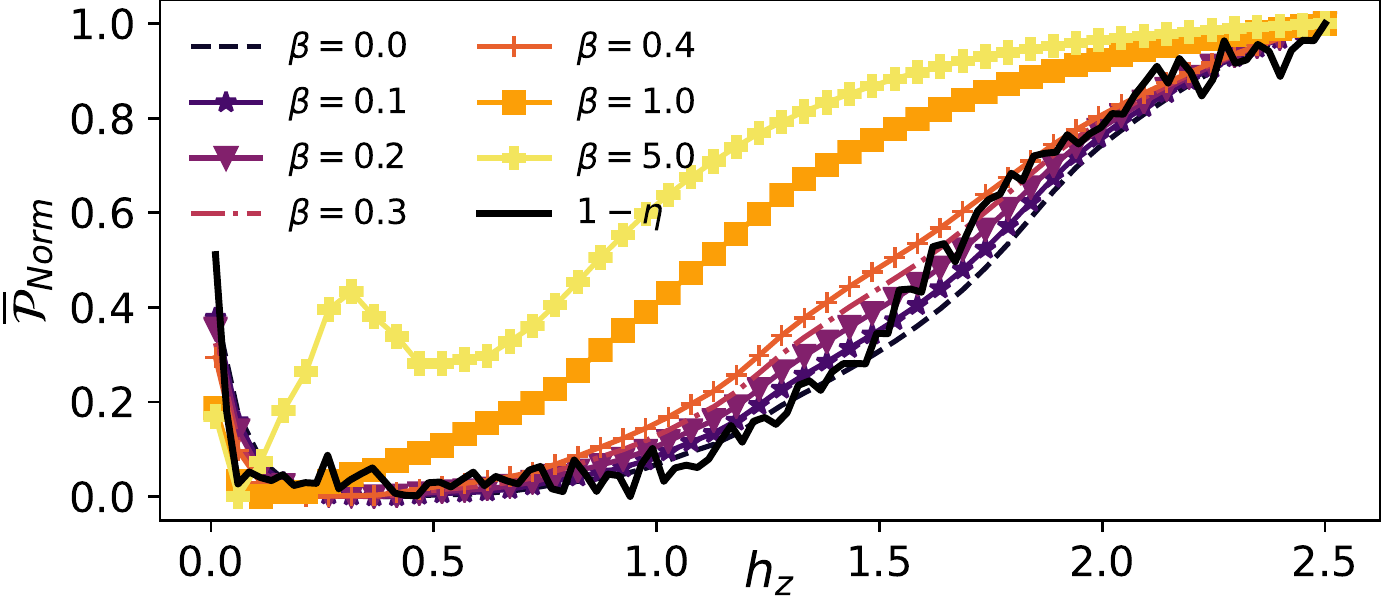}
\begin{footnotesize}
\caption{\nicon{$\mathcal{\overline{P}}_{Norm}$ as a function of $h_z$ for different environmental temperatures in a chain of $L=6$ spins, together with the chaos indicator $1-\eta$ for a long chain of $L=14$ spins. Parameters are set as $T=50$ (in units of $J^{-1}$), $h_x=1, J_k=1 \forall k$. 50 different realizations over initial random states of the probe were considered.}}
\label{beta_sup}
\end{footnotesize}
\end{center}
\end{figure}

\nicon{Other possibility is to consider different initial temperatures for the environmental \nicof{chain}. In the main text, we have restricted ourselves to a situation of infinite temperature, where all eigenstates and eigenenergies are equally contributing to the dynamics. This can be understood as if the probe dynamics is equally sampling the whole spectrum of the \nicof{chain}. On the contrary, if the temperature is lowered, eigenstates will contribute to the dynamics according to the Boltzmann factor, being the less energetic the ones that contribute more. This introduces an asymmetry in the task of sensing the chaoticity of the spectrum and therefore we should not expect the limit of low temperatures to give us consistent results, precisely because the spectrum is not being equally sampled by the probe. This is what we show in Fig. \ref{beta_sup}, where we plot $\mathcal{P}_{Norm}$ as a function of $h_z$ for different environmental temperatures. We can see that for sufficiently high temperatures, where almost all eigenstates are significantly contributing to the dynamics, chaoticity is quite well sampled. But for very low temperatures, since great part of the spectrum is being neglected, the dynamical signatures of chaos are not properly sensed. \nicof{In conclusion, the notion of purity loss, equilibration and quantum chaos are strictly related almost independently of the system size but only at the limit of high temperatures.}}  

\renewcommand{\figurename}{Figure}


\end{document}